# Survey on software testing techniques in cloud computing

V.Priyadharshini[#1], Dr. A. Malathi[*2]

[#] PhD Research Scholar
PG & Research Department of CS
Government Arts College (Autonomous)
Coimbatore – 18.
Tamil Nadu
India

[*] Assistant Professor
PG & Research Department of CS
Government Arts College (Autonomous)
Coimbatore – 18.
Tamil Nadu
India

[1] priyaphd13@gmail.com

*Abstract -* *Cloud computing is the next stage of the Internet evolution. It relies on sharing of resources to achieve coherence over a network. It is emerged as new computing standard that impacts several different research fields, including software testing. There are various software techniques used for testing application. It not only changes the way of obtaining computing resources but also changes the way of managing and delivering computing services, technologies, and solutions meanwhile, it causes new issues, challenges and needs in software testing. Software testing in cloud can reduce the need for hardware and software resources and offer a flexible and efficient alternative to the traditional software testing process. This paper provides an overview regarding trends, opportunities, challenges, issues and needs in cloud testing & cloud-based application.*

## I. INTRODUCTION

Software testing is performed to verify the completed software package functions according to the expectations defined by user. Testing allows developers to deliver the software that meets the expectations, prevents unexpected results, and improves long term maintenance of the application.

Cloud computing has gained significant attention in recent years as it changes the way of computation and providing the services to the customers whenever and wherever needed. It can be defined as set of hardware, networks, storage, services, and interfaces that combine to deliver aspects of computing as a service.

Cloud computing has four characteristics: Elasticity and Scalability, Multi-tenancy, Self-managed function capabilities, Billing and Service usage metering [1]. The purpose of this paper is to compare functional and non-functional testing in cloud computing environment.

## II. NEED FOR CLOUD COMPUTING

Cloud computing provides cost effective and flexible means through which scalable computing power and diverse services, application services are delivered as services to large-scale global users whenever and wherever needed.

## III. CLOUD TESTING ADVANTAGES

- Reduce cost by leveraging with computing resources in cloud – effectively use virtualized resources and shared cloud infrastructure [3].

- Advantage of on-demand test service to conduct large-scale and effective real-time online validation for internet based software clouds.

- Easily scalable cloud system infrastructure to test and evaluate performance and scalability.

## IV. PROBLEMS IN CLOUD COMPUTING

*Lack of control* – The IT infrastructure is outsourced to third party. How business maintain control over their data that lies beyond their boundaries.

*Security* – How far is the sensitive information that traverses through cloud is safe and secure?

*Privacy Concerns* – How businesses check their privacy of users and information maintained when using cloud.

*Data Integrity* – While using third party solutions to their problems, how do businesses assure their valuable data remains intact?

*Availability* – Computing solutions must be available to their customers to function effectively [4].

*Acceptability* – How businesses make sure that their third party solutions are planned for their use.

## V. SOFTWARE TESTING

It is a series of planned task that needs to be executed along with software development activities to ensure that a product is delivered without any errors. Conservative testing is done in two ways: 1) Functional Testing and 2) Non-functional Testing. The web based applications are powerful and have the ability to provide feature rich content to a wide audience spread across the globe [2]. These web applications are stored in remote server and accessed through the web browser. In order to produce the quality and secured web application, testing becomes the chief activity in web application development life cycle.

### A. Key Perception of cloud computing

It is particularly based on two key concepts. The first one is Service-Oriented Architecture (SOA), which is the delivery of an integrated and composed suite of functions to an end-user. The second key concept is virtualization [5]. Virtualization allows abstraction and isolation of lower level functionalities and hardware, which enables portability of higher level functions and sharing the physical resources.

### B. Characteristics

Rapid elasticity permits end users ease and rapid provision of new services and releases them, enabling them to pay for what they utilize and how much they use it. On-demand self-service is an appealing characteristic for consumers because a cloud computing provider pools its computing resources in order to serve multiple consumers by means of a multi-tenant provisioning model [7].

### C. Service delivery

Software as a Service (*SaaS*) delivery model describes software applications/services over cloud infrastructure for consumers. These applications are accessible from various platforms through an easy-to-use client interface such as a web browser.

Platform as a Service (*PaaS*) delivery model enables consumers to deploy their solutions to the cloud by means of platforms such as application servers and database services provided by the Cloud Platform Provider.

Infrastructure as a Service (*IaaS*) is the lowest level of service model in cloud delivery models. The consumers acquire computing services and can deploy their own custom configured systems [6].

### D. Cloud Deployment Model

*Public Cloud* - The cloud infrastructure is made available to the public or a large industry group and is owned by an organization selling cloud services.

*Private Cloud* - The cloud infrastructure operated solely for a single organization.

*Community Cloud* - The cloud infrastructure is share by several organizations and supports a specific community that has shared concerns.

*Hybrid Cloud* - The cloud infrastructure is a composition of two or more clouds.

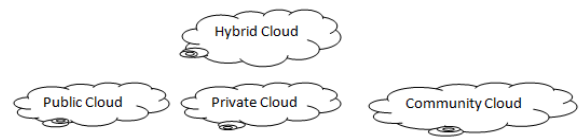

Fig. 1 Cloud Deployment Model

### E. Testing in cloud

It describes testing of applications that are specifically developed to run on a cloud platform. This fact entails that the application might be utilizing parallel computing features of cloud computing or it might be a multithreaded application [8]. Cloud service development and deployment, test task management, cloud infrastructure and storage, cloud applications domains are good examples of testing in cloud.

### F. Testing on cloud

It refers to the verification and validation of applications, environments and infrastructure that is available on demand. This ensures that applications, environments and infrastructure conform to the expectations of the cloud computing business model. For example, mobile and web applications are tested in multiple operating systems, multiple browser platforms and versions and different types of hardware to understand its performance in real-time.

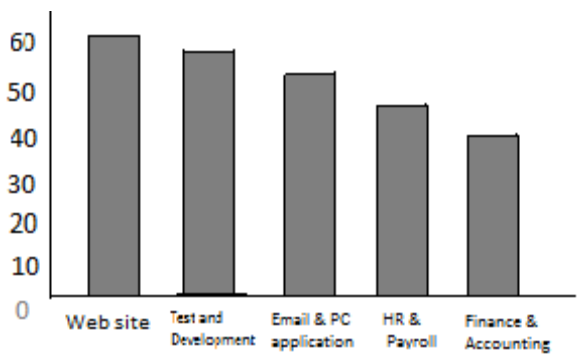

Fig. 2 Top application in cloud

Source : Cognizant Report

### VI. SERVICE MODEL OF CLOUD

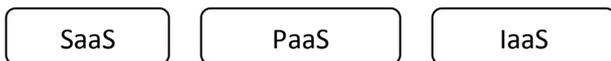

#### A Cloud computing in SaaS

Software as a Service (SaaS) is a type of cloud computing, which is a software delivery model. Software and its associated data are hosted centrally (typically in the (Internet) cloud) and are typically accessed by users using a thin client, normally using a web browser over the Internet [9]. Customers are not expected to buy software licenses or additional infrastructure equipment, and are expected to only pay monthly fees (also referred to as annuity payments) for using the software based on their usage.

#### B Cloud computing in PaaS

Cloud computing has evolved to include platforms for building and running custom applications, a concept known as "platform as a service" (or PaaS) PaaS can be considered as the next step in the SaaS model, where the on demand delivery is not just the specific item of software required, but the users' platform [11]. PaaS provides the entire infrastructure needed to run applications over the Internet. It is delivered in the same way as a utility like electricity or water. Users simply "tap in" and take what they need the complexities are hidden behind the scenes. And like any other utility, PaaS is based on a metering or subscription model, so users only pay for what they use again the delivery route in this model is the 'Cloud.

#### C Cloud computing in IaaS

The capability provided to the consumer is to provision processing, storage, networks, and other fundamental computing resources where the consumer is able to deploy and run arbitrary software, which can include operating systems and applications [10]. The consumer does not manage or control the underlying cloud physical infrastructure but has control over operating systems, storage, deployed applications, and possibly limited control of select networking components

### VII. TYPES OF TESTING IN CLOUD

The various types of testing performed in cloud are as follows,

*Unit Testing* – It is used to test a individual unit or group of related units. It can be described as testing of a function, module or object in isolation from rest of the program.

*Integration Testing* – It takes as input from unit test and groups them into single aggregate

*System Testing* – It is conducted on complete integrated system to evaluate system compliance with its specified requirements [12].

*Load Testing* – It describes an application involving creation of heavy user traffic and measuring its response.

*Acceptance Testing* – It is done to determine if the requirements of a specification is meeting user needs.

*Production Testing* – It is done when software is installed in client machine for real use [13].

## VIII. NEEDS AND CHALLENGES IN CLOUD COMPUTING

The current cloud technology does not have any supporting solutions that will help cloud engineers build a cost effective cloud test environment. A survey by Gao and others found that many of the published papers have discussed about performance testing and solutions.

*Scalability and performance testing* - they only focus on scalability evaluation metrics and frameworks for parallel and distributed systems. The current metrics, frameworks and solutions, does not support the features such as dynamic scalability [14].

*Regression testing* - Software challenges and bug fixing brings in regression testing issues and challenges. The on – demand cloud testing services should address the various issues and challenges.

*Adequate test models and criteria*- Test engineers should be provided with adequate test models and criteria that effectively support cloud testing.

*Continuous validation and regression testing solution* - If software has been changed due to bug fixing or for feature update, test engineers must provide automatic re- testing techniques which addresses the multi – tenancy feature of cloud.

*New automatic test solutions for cloud interoperability* - Test engineers should assure the interoperability quality of the cloud applications as both cloud and SaaS offers connectivity protocols and APIs [15].

### A Testing through Tools

Major technology vendors such as HP, Intel and Yahoo are presently collaborating to create huge cloud 'test beds' consisting of many thousands of processors working together as centers of excellence in Cloud Computing. The test beds allow users to test their cloud deployments at internet scale and also understand how their systems and software actually behave within the cloud [1]. Current test tool offerings by the likes of HP and IBM are ideal for non-functional and automated testing in a cloud environment. Well-established software such as HP's Quick Test Pro or IBM's Rational Robot can be used to full effect within a cloud environment to perform automated testing tasks such as regression test. Cloud Computing application and distributed architecture, as well as a good understanding of the tools available and their strengths and weakness for testing different types of cloud applications.

## IX. CONCLUSION

Functional testing acquires high usage of hardware and software to simulate user activity. While non-functional testing enables the measurement and association of the testing of non-functional attributes of software systems. Only a few advantages and few testing challenges of cloud computing have been identified. Testing is a periodic activity and new requirements need to be set up for each project.

ACKNOWLEDGEMENT

This paper is published in International Journal of Computer Trends and Technology